\documentclass[12pt]{article}
\usepackage{amsmath, amssymb}

\def\diag{\mathop{\rm diag}\nolimits}


\def\gtrsim{\mathrel{\mathpalette\vereq>}}
\makeatletter
\def\vereq#1#2{\lower3pt\vbox{\baselineskip1.5pt \lineskip1.5pt
\ialign{$\m@th#1\hfill##\hfil$\crcr#2\crcr\sim\crcr}}}
\makeatother
\newcounter{axn}

\def\diag{\mathop{\rm diag}\nolimits}


\newcommand{\del}{\partial}
\newcommand{\nn}{\nonumber}

\newcommand{\eff}{{\mrm {eff}}}

\newcommand{\bequ}{\begin{equation}}
\newcommand{\eequ}{\end{equation}}
\newcommand{\beqn}{\begin{eqnarray}}
\newcommand{\eeqn}{\end{eqnarray}}
\newcommand{\bctr}{\begin{center}}
\newcommand{\ectr}{\end{center}}
\newcommand{\Ls}{\left(}
\newcommand{\Rs}{\right)}
\newcommand{\Lm}{\left\{}
\newcommand{\Rm}{\right\}}

\newcommand{\LL}{\left.}
\newcommand{\RR}{\right.}
\newcommand{\hsp}[1]{\hspace {#1cm}}

\newcommand{\half}{{1\over2}}
\newcommand{\mrm}{\rm}

\def\PR#1#2#3{Phys. Rev. {\bf #1} (#3) #2 }
\def\PRL#1#2#3{Phys. Rev. Lett. {\bf #1} (#3) #2 }
\def\PL#1#2#3{Phys. Lett. {\bf #1} (#3) #2 }

\def\NP#1#2#3{Nucl. Phys. {\bf #1} (#3) #2 }

\def\PTP#1#2#3{Prog. Theor. Phys. {\bf #1} (#3)#2 }

\begin{document}
\begin{titlepage}
\begin{flushright}
KUNS-1945\\
OU-HET-502/2004
\end{flushright}
\begin{center}{\Large\bf Higgs mass in the gauge-Higgs unification}
\end{center}
\vspace{1cm}
\begin{center}
Naoyuki {Haba}$^{(a),}$
\footnote{E-mail: haba@ias.tokushima-u.ac.jp}
Kazunori {Takenaga}$^{(b),}$
\footnote{E-mail: takenaga@het.phys.sci.osaka-u.ac.jp}
Toshifumi {Yamashita}$^{(c),}$
\footnote{E-mail: yamasita@gauge.scphys.kyoto-u.ac.jp},
\end{center}
\vspace{0.2cm}
\begin{center}
${}^{(a)}$ {\it Institute of Theoretical Physics, University of
Tokushima, Tokushima 770-8502, Japan}
\\[0.2cm]
${}^{(b)}$ {\it Department of Physics, Osaka University, 
Toyonaka, Osaka 560-0043, Japan}
\\[0.2cm]
${}^{(c)}${\it Department of Physics, Kyoto University, Kyoto, 
606-8502, Japan}
\end{center}
\vspace{1cm}
\begin{abstract}
The gauge-Higgs unification theory identifies 
the zero mode of the extra dimensional component
of the gauge field as the usual Higgs doublet.
Since this degree of freedom is the Wilson
line phase, the Higgs does not have the 
mass term nor quartic coupling at the tree level.
Through quantum corrections, 
the Higgs can take a vacuum expectation value,
and its mass is induced. 
The radiatively induced mass tends to be small, 
although it can be lifted to ${\mathcal O}(100)$ GeV 
by introducing the ${\mathcal O}(10)$ numbers 
of bulk fields. Perturbation theory becomes unreliable
when a large number of bulk fields are introduced. We 
reanalyze the Higgs mass based on useful expansion formulae for
the effective potential  
and find that even a small number of bulk field
can have the suitable heavy Higgs mass. 
We show that a small (large) number of
bulk fields are enough (needed) when the SUSY breaking mass
is large (small). We also study the case of introducing 
the soft SUSY breaking 
scalar masses in addition to
the Scherk-Schwarz SUSY breaking and obtain the heavy Higgs
mass due to the effect of the scalar mass.
\end{abstract}
\end{titlepage}
\newpage
\section{Introduction}
There are much progress 
in the higher dimensional gauge theories. 
One of the most fascinating motivations 
for the higher dimensional gauge theory is that 
gauge and Higgs fields can be 
unified\cite{Manton:1979kb}-\cite{Hosotani:2004ka}.
The higher dimensional components of gauge fields
become scalar fields bellow the compactification scale, and  
these scalar fields are identified with 
the Higgs fields in the gauge-Higgs unification theory.
In fact, the adjoint Higgs fields can emerge through the $S^1$
compactification from 5D theory, while the Higgs doublet fields 
can appear through the orbifold compactification such as $S^1/Z_2$. 

In order to obtain the Higgs doublets from the gauge fields 
in higher dimensions, the gauge group must be lager than the 
standard model (SM) gauge group. The gauge symmetries are reduced 
by the orbifolding boundary conditions of the extra 
dimensions and can be broken further by the Hosotani 
mechanism \cite{YH}. The Higgs fields have only 
finite masses of order the compactification scale because 
the masses of the Higgs fields are forbidden by the higher dimensional 
gauge invariance. 

In the previous works\cite{HY1, HY2},
we have studied the possibility of the dynamical 
electro-weak symmetry breaking in two gauge-Higgs unified
models, $SU(3)_c \times SU(3)_W$ and $SU(6)$ models.
We calculated the one loop effective potential of Higgs doublets 
and analyze the vacuum structure of the models, and a similar 
analysis of the 6D gauge-Higgs unification model is 
studied in Ref.\cite{Hosotani:2004ka}. We found that the 
introduction of the appropriate numbers and representation 
of extra bulk fields are required for the desirable symmetry 
breaking. Since the Higgs is essentially the Wilson line 
degree of freedom, the mass term nor quartic coupling does 
not exist in the Higgs potential at the tree level. Through quantum 
corrections, the Higgs can develop a vacuum expectation value,
which means the dynamical electro-weak symmetry breaking 
is realized and accordingly its mass is induced. 
The induced Higgs mass tends to be small, less than the weak
scale, reflecting the nature of the Coleman-Weinberg 
mechanism\cite{coleman}.  

It is possible to lift the magnitude of the Higgs mass 
to ${\mathcal O}(100)$ GeV by introducing 
the ${\mathcal O}(10)$ numbers of bulk fields in Refs.\cite{HY1, HY2}.
The perturbation expansion is given by $g^2/16\pi^2$ 
$\times$ [bulk fields degrees of freedom]($g$: gauge coupling),
so that the analysis of the one loop effective potential
can not be reliable when there are a large number of bulk fields. 
Furthermore, it seems artificial to introduce a large number 
of extra fields. 

In this paper we reanalyze  and study the Higgs mass
analytically, not only numerically in the gauge-Higgs 
unification. Based on the expansion formulae for the 
effective potential, we show
that a small (large) number of bulk fields are needed when 
the SUSY breaking mass is large (small). 
We find that even a small number of bulk field
can make the Higgs mass suitably heavy. The expression for the 
Higgs mass is obtained by using the formula, which makes it clear 
that what mainly controls the magnitude of the Higgs mass. The
analyses made in this paper by using the expansion formula are 
applicable to the bulk field with an arbitrary 
representation under an arbitrary gauge group.

We also study the case that the soft SUSY breaking 
scalar masses exist in addition to the 
Scherk-Schwarz (SS) SUSY breaking\cite{SS}-\cite{SS4}. 
The soft scalar masse also plays the role to lift the Higgs
mass. And we show that also in this case 
a small (${\mathcal O}(1)$) number of extra bulk fields 
can realize the suitable electro-weak symmetry 
breaking and the Higgs mass of ${\mathcal O}(100)$ GeV. 

This paper is organized as follows. In section 2, we 
briefly overview the previous works.
In section 3 we will present useful expansion formulae for the 
effective potential, and by using them, we study the Higgs 
mass in the gauge-Higgs unified models. In section 4, we also 
study the Higgs mass for the case of existing soft SUSY breaking 
scalar masses. Section 5 devotes summary and discussion.

\section{Gauge-Higgs unification}

In this section we give a brief review 
of the dynamical electro-weak symmetry breaking 
in the $SU(3)_c \times SU(3)_W$ and $SU(6)$ models  
based on Refs.\cite{HY1, HY2}. In the gauge-Higgs unification 
models, the 5th dimensional coordinate is compactified 
on an $S^1/Z_2$ orbifold. The Higgs doublets are identified
as the zero modes of the extra dimensional component 
of the 5D gauge field, $A_5$. We 
check whether the dynamical electro-weak symmetry breaking is
possible or not by calculating one loop effective
potential of the Higgs doublets.  

Denoting $y$ as the coordinate of the 5th 
dimension, parity operator, $P$ ($P'$) are 
defined according to 
the $Z_2$ transformation, $y\rightarrow -y$
($\pi R -y\rightarrow \pi R+y$). In 
the $SU(3)_c \times SU(3)_W$ 
model\cite{Lim,Hall:2001zb,Burdman:2002se}, we 
take $P=P'=\diag(1,-1,-1)~(P=P'=I)$ in the 
base of $SU(3)_W (SU(3)_c)$.
Then, there appears Higgs doublet 
as the zero mode of $A_5$\footnote{Taking account
of the scalar degrees of freedom in the gauge 
super multiplet, we can easily show that there 
appear two Higgs doublets in the SUSY theory.},
\begin{equation}
\label{1a}
H=\sqrt{2\pi R} \; A_5. 
\end{equation}
The 4D gauge coupling constant is defined 
as $g_4 = g/\sqrt{2\pi R}$\footnote{
We should take $g_4 \gtrsim 1$ for 
the wall-localized kinetic terms
being the main part of the MSSM kinetic
terms\cite{HY1}. Hence, we take $g_4={\mathcal O}(1)$ in this paper.}. 
The vacuum expectation value (VEV) of
$A_5$ is parameterized by $a/(2gR){\bf E}_3$, where ${\bf E}_3$ is 
the $3\times3$ matrix having 1 at (1,3) and (3,1)
elements, while the other elements being zero\cite{HY1,HY2}.
The relation between the VEV and electro-weak scale is given by 
\begin{equation}
\label{3}
  {\sqrt{2\pi R}}\left<A_5^4\right>
 =\frac{a_0}{g_4R}=\; v\sim 246 \:{\rm GeV}.  
\end{equation}
Here the component gauge field $A_5^4$ is defined 
by $A_5=\sum_a A_5^a T^a$ through the 
generators $T^a$, where $T^4=\half{\bf E}_3$.
The compactification scale must be above 
 the weak scale, and when we take it as 
 a few TeV, for examples, $a_0$ should be a parameter 
 of ${\mathcal O}(10^{-1\sim -2})$.

Let us study SUSY theory with SS SUSY breaking. 
We define 
\beqn
J^{(+)}[a,\beta,n] &\equiv& 
   {1\over n^5}\Ls1-
      \cos(2\pi n\beta)\Rs 
      \cos(\pi n a), \nonumber \\
J^{(-)}[a,\beta,n] &\equiv& 
   {1\over n^5}\Ls1-
      \cos(2\pi n\beta)\Rs 
      \cos(\pi n (a-1)),\nonumber 
\label{J}
\eeqn
where $\beta (0\leq \beta \leq 0.5)$ parameterizes 
the magnitude of the SS SUSY breaking. Then, the soft mass 
parameters become ${\mathcal O}(\beta /R)$\cite{HY1,HY2}. 
The contribution of the gauge multiplet to the effective
potential is written as 
\begin{equation} 
V_{\eff}^{\rm gauge} = -2 C \sum_{n=1}^{\infty}
    \Ls J^{(+)}[2a,\beta,n]+2J^{(+)}[a,\beta,n] \Rs ,
\label{26}
\end{equation} 
where $C \equiv 3/(64\pi^7R^5)$. 
The VEV of $\sigma$, which forms 
the real part of scalar component
of $N=1$ chiral multiplet at low-energies, becomes
zero by calculation of the effective potential 
for $\langle\sigma\rangle$\cite{HTY}. The minimum of the
effective potential (\ref{26}) is located 
at $a_0=1$ (mod $2$), which means that the suitable electro-weak 
scale VEV, $(0<)a_0\ll 1$ and electro-weak symmetry breaking 
are not realized. Thus, for the desirable 
dynamical electro-weak symmetry breaking, one needs to 
introduce the extra bulk fields, which are 
$N_{fnd.}^{(\pm)}$ and $N_{adj.}^{(\pm)}$ species of
hypermultiplets of fundamental and adjoint 
representations, respectively. Here the index, ${(\pm)}$, denotes
the sign of the {\it intrinsic} parity of $P P'$ defined 
in Refs.\cite{HY1, HY2}.

The effective potential from the bulk fields is given by
\beqn
&&V_{\eff}^{\rm matter} = 2 C \sum_{n=1}^{\infty}\Lm
    N_{adj.}^{(+)}\Ls J^{(+)}[2a,\beta,n]
                    +2J^{(+)}[a,\beta,n] \Rs \RR \nn\\
&&
\hsp2 + N_{adj.}^{(-)}\Ls J^{(-)}[2a,\beta,n]
                    +2J^{(-)}[a,\beta,n] \Rs \nn\\
&& 
\hsp{1.95}  \LL 
 +  N_{fnd.}^{(+)}J^{(+)}[a,\beta,n]
         + N_{fnd.}^{(-)}J^{(-)}[a,\beta,n] \Rm.
\label{Vm}
\eeqn
Reference \cite{HY1} shows one 
example, $N_{adj.}^{(+)}=N_{adj.}^{(-)}=2$, 
$N_{fnd.}^{(-)}=4$, $N_{fnd.}^{(+)}=0$ with $\beta=0.1$
and $R^{-1}$ of order a few TeV, 
in which 
the suitable electro-weak symmetry breaking is realized by 
the small VEV, $a_0=0.047$. 
The Higgs mass is calculate by the
second derivative of the effective 
potential, $V_{\eff}\equiv C{\bar V}_{\eff}
=V_{\eff}^{\rm gauge}+V_{\eff}^{\rm matter}$
with respect to $a$ at the minimum, $a=a_0$, 
\begin{equation}
m_H\sim {{\sqrt{3}}\over {4\pi^3}}
\left({{\partial^2{\bar V}_{\eff}}\over
{\partial a^2}}\right)_{a=a_0}^{1/2}\times {{vg_4^2}\over a_0},
\end{equation}
where we have used (\ref{3}).
In this case Higgs mass is calculated as
\footnote{In Reference \cite{HY1}, we calculated it by using 
approximation formulae, and got slightly different value.}
\begin{equation} 
m_{H}^2 \sim \left({0.025~g_4 \over R}\right)^2
 \sim (118~g_4^2 \; {\rm GeV})^2,
\end{equation}
where $g_4 = {\mathcal O}(1)$, as explained above. 
The Higgs mass is likely to be smaller than 
the weak scale, 246 GeV (Eq.(\ref{3})) since 
it is zero at the tree level and is induced through the radiative 
corrections (Coleman-Weinberg mechanism).  

As for the $SU(6)$ model\cite{Hall:2001zb,Burdman:2002se},
we take the parities, $P=\diag(1,1,1,1,-1,-1)$
and $P'=\diag(1,-1,-1,-1,-1,-1)$, which 
induces Higgs doublet in $A_5$ as the zero mode. 
The VEV of $A_5$ is written 
as $a/(2gR){\bf E}_6$, where ${\bf E}_6$ is 
the $6\times6$ matrix having 1 at (1,6) and (6,1)
elements while the other elements being zero\cite{HY1,HY2}.
The gauge part of the effective potential is given by
\begin{equation} 
V_{\eff}^{\rm gauge} = -2 C \sum_{n=1}^{\infty}
    \Ls J^{(+)}[2a,\beta,n]+2J^{(+)}[a,\beta,n] 
      +6J^{(-)}[a,\beta,n] \Rs .
\label{VSU6}
\end{equation}
As in the $SU(3)_c \times SU(3)_W$ model, the suitable 
symmetry breaking can not be realized only by the 
gauge sector. This situation can be changed by 
introducing the extra bulk fields, which induce 
the effective potential, 
\beqn
V_{\eff}^{\rm matter} &=& 2 C \sum_{n=1}^{\infty}\Lm
    N_{adj.}^{(+)}\Ls J^{(+)}[2a,\beta,n]
                    +2J^{(+)}[a,\beta,n] 
                    +6J^{(-)}[a,\beta,n] \Rs \RR\nn\\
  &&\qquad+ N_{adj.}^{(-)}\Ls J^{(-)}[2a,\beta,n]
                    +2J^{(-)}[a,\beta,n] 
                    +6J^{(+)}[a,\beta,n] \Rs \nn\\
  &&\LL\qquad
  + N_{fnd.}^{(+)}J^{(+)}[a,\beta,n]
         + N_{fnd.}^{(-)}J^{(-)}[a,\beta,n] \Rm .
\label{VSU6-m}
\eeqn
We show one example of the suitable symmetry 
breaking in Ref.\cite{HY1}, which is the case of
$N_{adj.}^{(+)}=2$, $N_{fnd.}^{(-)}=10$, 
$N_{adj.}^{(-)}=N_{fnd.}^{(+)}=0$ with $\beta=0.1$
and $R^{-1}$ of order a few TeV.
In this case, the minimum exists 
at $a_0=0.047$, and the Higgs mass squared is 
calculated as
\begin{equation} 
m_{H}^2 \sim \left({0.024~g_4 \over R}\right)^2
 \sim (120~g_4^2 \; {\rm GeV})^2.
\end{equation}

In the above two examples ${\mathcal O}(10)$ numbers of
bulk fields are required for the suitable symmetry breaking
and Higgs mass. Naively, this situation seems inevitable
in the gauge-Higgs unification theory since Higgs doublets
are originally Wilson line phases and do not have the quartic
couplings nor mass terms in the tree level. 

In the next section we obtain the effective mass term 
and quartic coupling by expanding the cosine functions 
with respect to $a$ in the effective potential
and study the condition for the suitable 
symmetry breaking and Higgs mass. 
We will check whether a large numbers
of extra fields are really needed or not.

\section{Higgs mass}

The Higgs mass is defined by the 2nd derivative of the effective 
potential. Here we concentrate on the mass term and quartic
coupling in the radiatively induced effective potential. We 
comment on higher order terms in the last section.
The effective potentials in the previous section are written 
by the linear combinations of $J^{(\pm)}[a,\beta,n]$ 
and $J^{(\pm)}[2a,\beta,n]$. Effective potentials are generally 
the linear combination of $J^{(\pm)}[ma,\beta,n]$, $(m: {\rm integer})$,
({\it see for examples,} Refs.\cite{HY1, HY2, Haba:2004xx}).  
Here we study the contribution from the fundamental representation 
bulk fields for simplicity. Although the gauge sector and adjoint
representation bulk fields also induce $J^{(\pm)}[2a,\beta,n]$ terms
(higher representations can induce $J^{(\pm)}[ma,\beta,n]$ in general),
these contributions can be incorporated straightforwardly. 

By using approximations for small $x$, we 
obtain, up to ${\mathcal O}(x^8)$, that
\beqn
&& \sum_{n=1}^{\infty}\frac{\cos(nx)}{n^5}\sim
    \zeta_R(5)-{{\zeta_R(3)}\over 2}x^2-\frac{x^4}{48}\ln\Ls x^2\Rs
      +\frac{25}{288}x^4+\frac{x^6}{8640}+\frac{x^8}{4838400}, 
\label{exp1}\\
&& \sum_{n=1}^{\infty}\frac{\cos(n(x-\pi))}{n^5}\sim
    -\frac{15}{16}\zeta_R(5)+\frac38\zeta_R(3)x^2-\frac{x^4}{24}\ln2
      +\frac{x^6}{2880}+
     \frac{x^8}{322560}.
\label{exp2}
\eeqn
Let us note that these formulae are also applicable to 
the bulk field with the higher representations under 
the gauge group. Then, we can show that
\beqn
\label{90}
&& \sum_{n=1}^{\infty}
 J^{(+)}[\frac{a}{\pi},\frac{\beta}{\pi},n]
 \sim
    \frac{a^2}{288}\Ls 25a^2 - 432\beta^2 
       - 
         6a^2\ln\Ls\frac{a^2}{4\beta^2}\Rs 
       + 144\beta^2\ln(4\beta^2)\Rs, \\
&& \sum_{n=1}^{\infty}
J^{(-)}[\frac{a}{\pi},\frac{\beta}{\pi},n]
\sim
    -\frac{\beta^2}{48}\Ls a^4 - 48a^2\ln2\Rs. 
\label{100}
\eeqn
for $a\ll \beta$\footnote{In the usual scenario, 
$a < \beta$ should be satisfied since the SUSY breaking mass, 
${\mathcal O}(\beta /R)$ must be larger than the electro-weak scale, 
${\mathcal O}(a/R)$.}.
Then, the coefficients of 
$a^2$ and $a^4$ in Eq.(\ref{90}) are roughly
given by 
\begin{equation}
\label{9}
-{\beta^2\over288}(432-144 \ln (4\beta^2))\;\; (<0)
\;\;\;{\rm and}\;\;\;
{25-6\ln{(a^2_0/4\beta^2)}\over288}\;\; (>0),
\end{equation}
respectively. On the other hand,
coefficients of $a^2$ and $a^4$ in Eq.(\ref{100}) are
\begin{equation}
\label{10}
 \beta^2\ln 2\;\; (>0),
\;\;\;\;{\rm and}\;\;\;\;\;
-\beta^2/48\;\; (<0),
\end{equation}
respectively. 

For realizing the suitable heavy Higgs 
mass, the quartic coupling should be large and positive.
On the other hand, the VEV (W and Z boson masses) should be 
maintained small $(a_0\ll 1)$ comparing to the compactification 
scale. For this purpose, large negative 
contribution in the first term in Eq.(\ref{9}) must be almost 
cancelled by introducing $N_{fnd.}^{(-)}={\mathcal O}(\ln 4\beta^2)$ 
numbers bulk fields acting on the first term in Eq.(\ref{10}).
This means that the less (more) bulk fields are 
needed when $\beta$ becomes large (small). 
Eqs.(\ref{9}) and (\ref{10}) shows that even in the case of this 
cancellation, the coefficient of $a^4$ is still positive and 
large enough when $a_0\ll\beta$.
Thus, the heaviness of Higgs mass is mainly controlled by 
the factor $-\ln (a^2_0/\beta^2)$ in the effective quartic coupling,
which implies that the smaller (larger) $a^2_0/\beta^2$ becomes, 
the larger (smaller) the Higgs mass becomes. A typical numerical 
example for realizing this in the $SU(3)_c\times SU(3)_W$ model 
is given in the table.
\[
\begin{array}{|cccc|ccc|}
\hline
 N_{adj.}^{(+)} & N_{adj.}^{(-)} &
N_{fnd.}^{(+)} & N_{fnd.}^{(-)} & \beta & a_0 & m_H/g_4^2~\mbox{(GeV)} 
\\ \hline\hline
 2 & 2 & 0 & 2 & 0.10 & 
 0.0891 & 95 \\ \hline
 2 & 2 & 0 & 2 & 0.13 & 
 0.0574 & 117 \\ \hline
 2 & 2 & 0 & 2 & 0.14 & 
 0.0379 & 130 \\ \hline
\end{array}
\]
%
%
The observation given above becomes clearer  
if we apply (\ref{90}) and (\ref{100}) to the effective 
potential, $V_{\eff}$ and take the order up 
to $O(x^4)$. Then, the Higgs mass for the $SU(3)_c \times
SU(3)_W$ model is calculated as 
\begin{equation}
{m_H\over g_4^2}\simeq v~{\sqrt{3}\over {4\pi}}
\sqrt{4B~{\rm ln}\left({{a_0^2}\over{4\beta^2}}\right)+{\rm const.}},
\label{higgsmass}
\end{equation}
where 
\begin{equation}
B\equiv {{-1}\over
{24}}\left(18(N_{adj.}^{(+)}-1)+N_{fnd.}^{(+)}\right),
\label{coeff}
\end{equation}
and the constant term depends on $\beta$ and the number of
flavors. Equation (\ref{coeff}) shows that a few 
adjoint bulk fields are enough
and essential for the large quartic coupling. The contribution 
from the adjoint bulk field overcome the loop factor $\sim
1/4\pi$ to enhance the magnitude of the Higgs mass. 
Let us note that the dependence of the Higgs mass 
on the supersymmetry breaking parameter is 
logarithmic, as expected. 

If we tune the values of $\beta$ to $\beta_c\simeq 
0.14865\cdots$ at which the coefficient of $a^2$ vanishes at 
one-loop level, the smaller VEV $a_0$ is realized within the
validity of perturbation theory, and the scale
$R$ should be smaller due to Eq.(\ref{3}).
In this sence two examples below are different
 theories from each other, since they should have 
 different initial setup 
 of the value, $R$.
The magnitude of the Higgs mass is enhanced because of the 
large ${\rm ln}\left(a_0^2/\beta^2\right)$ as shown in the table
for the $SU(3)_c\times SU(3)_W$ model with 
$(N_{adj.}^{(+)}, N_{adj.}^{(-)}, 
N_{fnd.}^{(+)}, N_{fnd.}^{(-)})=(2, 2, 0, 2)$ \footnote{Note that
too small $a_0$ induces a very large log factor that spoils the
validity of perturbation theory.}. 
\[
\begin{array}{|cccc|ccc|}
\hline
 N_{adj.}^{(+)} & N_{adj.}^{(-)} &
N_{fnd.}^{(+)} & N_{fnd.}^{(-)} & \beta & a_0 & m_H/g_4^2~\mbox{(GeV)}
\\ \hline\hline
 2 & 2 & 0 & 2 & 0.1486  &  0.0023   & 191 \\ \hline
 2 & 2 & 0 & 2 & 0.14865 &  0.0009   & 208 \\ \hline
\end{array}
\]
Two loop contributions become dominant in the coefficient
of $a^2$ if we tune $\beta$ close to $\beta_c$. The 
effect of the two loop, however, is 
almost absorbed into the values of $\beta$ 
by adjusting $\beta$ at one-loop level as long as the
perturbation is valid. If one chooses $\beta$ such that 
the coefficient of $a^2$ taken account of higher loops 
almost vanishes, one expects very small values of $a_0$, so that
the magnitude of the Higgs mass becomes larger 
than the values obtained in the above table.

Let us comment on the heavy Higgs mass in non-SUSY gauge models.
We can see from Eqs.(\ref{exp1}) and (\ref{exp2}) that
it is possible to cancel the $a^2$ terms between the matter with 
even parity and the one with odd parity, keeping the positive and 
large quartic coupling, by an appropriate choice of 
the matter content. In this case, we have the similar situation
with the SUSY case studied above and expect the desirable 
size of the Higgs mass. In fact, the non-SUSY model with the 
appropriate matter content, which 
realizes the suitable dynamical electro-weak symmetry 
breaking, is presented in Ref.\cite{HY1}.    
\section{SUSY gauge-Higgs with bulk mass}
In this section we show another example
for realizing the dynamical electro-weak symmetry breaking
with the small number of extra bulk fields.
We introduce explicit soft SUSY breaking 
scalar mass in addition to the SS parameter for the 
bulk superfields\footnote{The effective potentials 
and vacuum structures with soft scalar masses 
on $S^1$ have been studied in 
Refs.\cite{Takenaga:2003dp, Haba:2004zx}.}.
In this paper we do not introduce the soft gaugino masses
because the mass terms are odd under the $Z_2$ operation. 

Let us study the $SU(3)_c \times SU(3)_W$ model at first. 
The contribution of the gauge multiplet to the 
effective potential is the same as Eq.(\ref{26}). 
We introduce the soft SUSY breaking mass, $m$ for 
the bulk hypermultiplets and define a dimensionless 
parameter, $z \equiv m R \; (<1)$. We denote 
\beqn
\label{zeff}
I^{(+)}[a,\beta,z,n] &\equiv& 
   {1\over n^5}\Ls1-\Ls1+2\pi zn+{{(2\pi zn)^2}\over 3}\Rs e^{-2\pi zn}
      \cos(2\pi n\beta)\Rs \nonumber\\
&\times & \cos(\pi n a), \\ 
I^{(-)}[a,\beta,z,n] &\equiv& 
   {1\over n^5}\Ls1-\Ls1+2\pi zn+{{(2\pi zn)^2}\over 3}\Rs e^{-2\pi zn}
      \cos(2\pi n\beta)\Rs \nonumber\\ 
    &\times &  \cos(\pi n (a-1)), 
\eeqn
in which $I^{(\pm)}[a,\beta,z,n]$ is reduced 
to $J^{(\pm)}[a,\beta,n]$ in the limit 
of $z \rightarrow 0$ ($m \rightarrow 0$). The contribution 
of the matter hypermultiplet to the effective potential 
is given by
\beqn
V_{\eff}^{\rm matter} &=& 2 C \sum_{n=1}^{\infty}\Lm
    N_{adj.}^{(+)}\Ls I^{(+)}[2a,\beta,z_{adj.}^{(+)},n]
                    +2I^{(+)}[a,\beta,z_{adj.}^{(+)},n] \Rs \RR\nn\\
  &&\qquad+ N_{adj.}^{(-)}\Ls I^{(-)}[2a,\beta,z_{adj.}^{(-)},n]
                    +2I^{(-)}[a,\beta,z_{adj.}^{(-)},n] \Rs \nn\\
  &&\LL\qquad+ N_{fnd.}^{(+)}I^{(+)}[a,\beta,z_{fnd.}^{(+)},n]
         + N_{fnd.}^{(-)}I^{(-)}[a,\beta,z_{fnd.}^{(-)},n] \Rm ,
\label{Vm2}
\eeqn
where $z_{rep.}^{(\pm)}$ stands for the explicit soft 
mass defined by $z_{rep.}^{(\pm)}\equiv m_{rep.}^{(\pm)} R \; (<1)$
for each representation field. Eq.(\ref{Vm2})
becomes Eq.(\ref{Vm}) in the limit of the vanishing soft 
scalar mass, $m\rightarrow 0$. 

We find some examples of extra matter contents 
and SUSY breaking parameters, for which the suitable VEV 
and Higgs mass are realized, and we 
summarize them in the following table.
\[
\begin{array}{|c|cccc|ccccc||cc|}
\hline
 &N_{adj.}^{(+)} & N_{adj.}^{(-)} &
N_{fnd.}^{(+)} & N_{fnd.}^{(-)} & \beta &
z_{adj.}^{(+)} & z_{adj.}^{(-)} &
z_{fnd.}^{(+)} & z_{fnd.}^{(-)} &
a_0 & m_H/g_4^2 \\ \hline\hline
(1)& 2 & 3 & 0 & 4 & 0.05 & 0.01 & 0.01 & \mbox{-} & 0.045 &
 0.0040 & 164 \\ \hline
(2)& 2 & 4 & 2 & 6 & 0.05 & 0 & 0 & 0.05 & 0.05 & 
 0.0037 & 176 \\ \hline
(3)& 2 & 4 & 0 & 6 & 0.025 & 0.025 & 0.025 & \mbox{-} & 0.025 & 
 0.0066 & 129 \\ \hline
(4)& 2 & 1 & 0 & 2 & 0.1 & 0.1 & 0.1 & \mbox{-} & 1 & 
 0.0097 & 150 \\ \hline
(5)& 1 & 1 & 0 & 2 & 0.01 & 1 & 1 & \mbox{-} & 1 & 
 0.0196 & 125 \\ \hline
(6)& 2 & 2 & 0 & 2 & 0.14 & 0 & 0 & \mbox{-} & 0 & 
 0.0379 & 130 \\ \hline
\end{array}
\]
The Higgs mass $m_H/g_4^2$ is measured in GeV unit.
This table shows that even small number of extra bulk 
fields can realize the suitable dynamical electro-weak 
symmetry breaking with the heavy Higgs mass. 
The effect of the bulk masses increases not only the degrees 
of freedom of parameter space, but also induces a similar effect
of large $\beta$, which is necessary for the symmetry breaking
with a small number bulk fields, as explained in the last
section. We show an example in which one can see the
enhancement of the magnitude of the Higgs mass due 
to the existence of the bulk mass in 
the table, where the Higgs 
mass $m_H/g_4^2$ is measured in GeV unit.
\[
\begin{array}{|cccc|ccccc||cc|}
\hline
N_{adj.}^{(+)} & N_{adj.}^{(-)} &
N_{fnd.}^{(+)} & N_{fnd.}^{(-)} & \beta &
z_{adj.}^{(+)} & z_{adj.}^{(-)} &
z_{fnd.}^{(+)} & z_{fnd.}^{(-)} &
a_0 & m_H/g_4^2 \\ \hline\hline
2 & 1 & 0 & 2 & 0.1 & 0 & 0 & \mbox{-} & 0 & 
 0.2362 & 42 \\ \hline
2 & 1 & 0 & 2 & 0.1 & 0.1 & 0.1 & \mbox{-} & 1 & 
 0.0097 & 150 \\ \hline
\end{array}
\]

%
%
%

Next we study the case of $SU(6)$ model. 
The contribution of the gauge multiplet 
to the effective potential is the same as 
Eq.(\ref{VSU6}). On the other hand, 
the contribution of the matter hypermultiplet 
to the effective potential is given by
\beqn
V_{\eff}^{\rm matter} &=& 2 C \sum_{n=1}^{\infty}\Lm
    N_{adj.}^{(+)}\Ls I^{(+)}[2a,\beta,z_{adj.}^{(+)},n]
                    +2I^{(+)}[a,\beta,z_{adj.}^{(+)},n] 
                    +6I^{(-)}[a,\beta,z_{adj.}^{(+)},n] \Rs \RR\nn\\
  &&\qquad+ N_{adj.}^{(-)}\Ls I^{(-)}[2a,\beta,z_{adj.}^{(-)},n]
                    +2I^{(-)}[a,\beta,z_{adj.}^{(-)},n] 
                    +6I^{(+)}[a,\beta,z_{adj.}^{(-)},n] \Rs \nn\\
  &&\LL\qquad+ N_{fnd.}^{(+)}I^{(+)}[a,\beta,z_{fnd.}^{(+)},n]
         + N_{fnd.}^{(-)}I^{(-)}[a,\beta,z_{fnd.}^{(-)},n] \Rm ,
\eeqn
which becomes Eq.(\ref{VSU6-m}) in the zero limit of 
explicit soft scalar masses. Some examples for realizing 
the suitable dynamical electro-weak symmetry breaking 
are shown in the following table.
%
\[
\begin{array}{|c|cccc|ccccc||cc|}
\hline
 & N_{adj.}^{(+)} & N_{adj.}^{(-)} &
N_{fnd.}^{(+)} & N_{fnd.}^{(-)} & \beta &
z_{adj.}^{(+)} & z_{adj.}^{(-)} &
z_{fnd.}^{(+)} & z_{fnd.}^{(-)} &
a_0 & m_H/g_4^2 \\ \hline\hline
(7)& 2 & 0 & 0 & 10 &  0.1 & 0.05 & \mbox{-} & \mbox{-} & 0.05 & 
  0.0207 & 139 \\ \hline
(8)& 2 & 0 & 0 &  6 & 0.15 & 0.1 & \mbox{-} & \mbox{-} &  0.1 & 
  0.0268 & 139 \\ \hline
(9)& 2 & 0 & 0 & 16 & 0.04 &  0 & \mbox{-} & \mbox{-} &  0.03 &  
  0.0021 & 173 \\ \hline
(10)& 2 & 0 & 0 &  4 & 0.07 & 0.5 & \mbox{-} & \mbox{-} &  0.5 &  
  0.0366 & 138 \\ \hline
(11)& 2 & 0 & 0 &  2 & 0.32 &  0 & \mbox{-} & \mbox{-} &   0 &  
  0.0594 & 135 \\ \hline
\end{array}
\]
\section{Summary and discussions}

We have studied the Higgs mass in the gauge-Higgs 
unification theory. Since the Higgs doublet corresponds to
the Wilson line phases, it does not have the mass 
term nor quartic coupling at the tree level.
Through the quantum corrections, the Higgs can take a VEV, 
and its mass is induced. The radiatively induced 
mass, however, tends to be small, so we lift 
it to ${\mathcal O}(100)$ GeV by introducing 
the ${\mathcal O}(10)$ numbers of bulk fields
in the previous works. The perturbation theory can 
not be reliable when there are a large number of bulk fields. 

In this paper we have reanalyzed the Higgs mass
and have found that even a small number of bulk field
can have the suitable heavy Higgs mass, accompanying the 
desirable electro-weak symmetry breaking. The expansion formulae
for the effective potential are useful to discuss and study
analytically the Higgs mass. And we have shown that a 
small (large) number of
bulk fields are enough (needed) when the SUSY breaking mass
is large (small). The Higgs mass has the logarithmic 
dependence on the supersymmetry breaking parameter of the
Scherk-Schwarz mechanism. 
The fine tuning of $\beta$ yields
smaller VEV, $a_0$, and accordingly enhances the magnitude of the
Higgs mass. 
The analyses in the paper can be
applied to the bulk field with an arbitrary representation 
under an arbitrary gauge group. We have also studied the 
case of introducing the soft SUSY breaking scalar 
masses in addition to the SS SUSY breaking. In this 
case the suitable electro-weak 
symmetry breaking and the ${\mathcal O}(100)$ GeV  Higgs mass 
can also be realized by ${\mathcal O}(1)$ numbers of bulk fields.

Finally, we would like to discuss the higher order operators 
of Higgs self interactions. We see that the effective 
potential contains $a^n$ interactions by the expansion of 
the cosine function, which implies $a^n =(g_4 RH)^n$ from 
Eqs.(\ref{1a}) and (\ref{3}). 
When $g_4 R$ is of order a few TeV, 
 higher order operators, $H^n$ $(n\geq 6)$ have the 
 dimensionful suppression of order a few TeV.
This means that the contributions from the higher order 
operators are not so significant. 

In connection with new physics expected in the scenario of the
gauge-Higgs unification, it may be interesting to 
comment on the effective 3-point self coupling of $H$ in the 
models. The coupling is important for the search of 
the new physics in the future linear colliders\cite{Kanemura:2002vm}. 
The coupling of the effective $\lambda H^3$ interaction is 
given by $\lambda \equiv {3g_4^3\over 32\pi^6 R}
\left.{\del^3(V/C)\over \del a^3}\right|_{a_0}$, 
and the deviation from the tree level SM 
coupling, $\lambda_{SM}=3m_h^2/v$, is 
estimated by $\Delta \lambda=(\lambda
-\lambda_{SM})/\lambda_{SM}$ \cite{Kanemura:2002vm}.
The value of $\Delta \lambda$ becomes $-17.4$\% for the example 
of $SU(3)_c \times SU(3)_W$ model
($N_{adj.}^{(+)}=N_{adj.}^{(-)}=2$, 
$N_{fnd.}^{(-)}=4$, $N_{fnd.}^{(+)}=0$ with $\beta=0.1$) and 
$-16.6$\% for $SU(6)$ model ($N_{adj.}^{(+)}=2$, 
$N_{fnd.}^{(-)}=10$, $N_{adj.}^{(-)}=N_{fnd.}^{(+)}=0$ 
with $\beta=0.1$).
As for the examples of section 4, we show them in the following table.
{\small 
\[
\begin{array}{|c|ccccccccccc|}
\hline
 & (1) & (2) & (3) & (4) & (5) & (6) & (7) & (8) & (9) & (10) & (11) \\ \hline
\Delta \lambda (\%) & 
 -8.6 & -8.3 & -14.0 & -10.2 & -3.1 & -13.7 & -12.0 & -12.0 & -7.6 &
 -11.2 & -12.7 \\ \hline
\end{array}
\]
}
The effective 3-point self couplings tend to be small
comparing to that of the SM.
We should notice again that  
the Higgs field in our model has VEV 
in $A_5$ not $\sigma$. The VEV of $A_5$ should be 
distinguished from that of 
$\sigma$ in the dynamically induced effective
potential in the gauge-Higgs unification theory\cite{HTY}. 
Since $H$ is the field of the $D$-flat direction,
which is massless at tree level, it corresponds to 
the lighter Higgs scalar in the MSSM, $h^0$.      
Since $h^0$ becomes the SM-like Higgs in the 
large soft SUSY breaking masses, 
we have compared the effective 3-point self coupling to 
the SM one in the above estimation of $\Delta \lambda$. 
Other masses of Higgs eigenstates, charged 
Higgs, neutral scalar, and heavier pseudo-scalar, can be
calculated by the effective potential of 
these directions\cite{HTY2}.

\vskip 1.5cm

\leftline{\bf Acknowledgements}

We would like to thank M. Tanabashi and Y. Okada
for useful discussions which become one of the motivations
of this work. We would like to thank N. Okada for 
a lot of useful and helpful discussions. 
K.T. would thank the colleagues in Osaka University and gives special
thanks to the professor Y. Hosotani for valuable discussion.
K.T. is supported by the $21$st Century COE Program at Osaka 
University. T.Y. would like to thank the Japan Society
for the Promotion of Science for financial support. 
N.H. is supported in part by Scientific Grants from 
the Ministry of Education and Science, 
Grant No.\ 14740164, No.\ 16028214, and No.\ 16540258.

\vspace{1cm}

\def\jnl#1#2#3#4{{#1}{\bf #2} (#4) #3}

\def\Zphys{{\em Z.\ Phys.} }
\def\jssc{{\em J.\ Solid State Chem.\ }}
\def\jpsJ{{\em J.\ Phys.\ Soc.\ Japan }}
\def\ptps{{\em Prog.\ Theoret.\ Phys.\ Suppl.\ }}
\def\PTP{{\em Prog.\ Theoret.\ Phys.\  }}

\def\JMP{{\em J. Math.\ Phys.} }
\def\NPB{{\em Nucl.\ Phys.} B}
\def\NP{{\em Nucl.\ Phys.} }
\def\PLB{{\em Phys.\ Lett.} B}
\def\PL{{\em Phys.\ Lett.} }
\def\PRL{\em Phys.\ Rev.\ Lett. }
\def\PRB{{\em Phys.\ Rev.} B}
\def\PRD{{\em Phys.\ Rev.} D}
\def\PRe{{\em Phys.\ Rep.} }
\def\AP{{\em Ann.\ Phys.\ (N.Y.)} }
\def\RMP{{\
em Rev.\ Mod.\ Phys.} }
\def\ZPC{{\em Z.\ Phys.} C}
\def\SCI{\em Science}
\def\CMP{\em Comm.\ Math.\ Phys. }
\def\MPLA{{\em Mod.\ Phys.\ Lett.} A}
\def\IJMPA{{\em Int.\ J.\ Mod.\ Phys.} A}
\def\IJMPB{{\em Int.\ J.\ Mod.\ Phys.} B}
\def\EPJC{{\em Eur.\ Phys.\ J.} C}
\def\PR{{\em Phys.\ Rev.} }
\def\JHEP{{\em JHEP} }
\def\cmp{{\em Com.\ Math.\ Phys.}}
\def\JPA{{\em J.\  Phys.} A}
\def\CQG{\em Class.\ Quant.\ Grav. }
\def\ATMP{{\em Adv.\ Theoret.\ Math.\ Phys.} }
\def\ibid{{\em ibid.} }

\leftline{\bf References}

\renewenvironment{thebibliography}[1]
         {\begin{list}{[$\,$\arabic{enumi}$\,$]}  
         {\usecounter{enumi}\setlength{\parsep}{0pt}
          \setlength{\itemsep}{0pt}  \renewcommand{\baselinestretch}{1.2}
          \settowidth
         {\labelwidth}{#1 ~ ~}\sloppy}}{\end{list}}

\end{document}